%% file: main_v3.tex
\documentclass[aps,twocolumn,superscriptaddress,floatfix,nofootinbib,showpacs,altaffilletter]{revtex4-2}

\usepackage{epsfig,graphicx,amssymb,amsmath,wasysym,graphicx}

\usepackage{amsmath}
\usepackage{amsfonts}
\usepackage{amssymb}
\usepackage{graphicx}
\usepackage{bm}
\usepackage{subfigure}
\usepackage{url}
\usepackage[hyperindex]{hyperref}
\usepackage{color}
\usepackage[ddmmyy,24hr]{datetime}
\usepackage{bigdelim}
\usepackage{booktabs}
\usepackage{dcolumn}
\usepackage{multirow}
\usepackage{subfigure}
\usepackage{cancel}
\usepackage{stackrel}
\usepackage{paralist}
\usepackage{xspace}
\usepackage{slashed}
\usepackage{cancel}
\usepackage{todonotes}
\usepackage{enumerate}
\usepackage{float}
\usepackage{fullpage}
\usepackage{ulem}
\usepackage{physics}

\newcommand{\nua}[1]{\ensuremath{\rlap{\kern-2.5pt\ensuremath{\overset{\scriptscriptstyle(-)}{\phantom{\nu}}}}{\ensuremath{{\nu}_{#1}}}}}

\usepackage{enumitem}

\newcommand{\be}{\begin{equation}}
\newcommand{\ee}{\end{equation}}
\newcommand{\ba}{\begin{array}}
\newcommand{\ea}{\end{array}}

\begin{document}

\title{New constraint on neutrino magnetic moment and neutrino millicharge from LUX-ZEPLIN dark matter search results}

\author{M. Atzori Corona}
\email{mattia.atzori.corona@ca.infn.it}
\affiliation{Dipartimento di Fisica, Universit\`{a} degli Studi di Cagliari,
	Complesso Universitario di Monserrato - S.P. per Sestu Km 0.700,
	09042 Monserrato (Cagliari), Italy}
\affiliation{Istituto Nazionale di Fisica Nucleare (INFN), Sezione di Cagliari,
	Complesso Universitario di Monserrato - S.P. per Sestu Km 0.700,
	09042 Monserrato (Cagliari), Italy}
	
\author{W. M. Bonivento}
\email{walter.bonivento@ca.infn.it}
\affiliation{Istituto Nazionale di Fisica Nucleare (INFN), Sezione di Cagliari,
	Complesso Universitario di Monserrato - S.P. per Sestu Km 0.700,
	09042 Monserrato (Cagliari), Italy}

\author{M. Cadeddu}
\email{matteo.cadeddu@ca.infn.it}
\affiliation{Istituto Nazionale di Fisica Nucleare (INFN), Sezione di Cagliari,
	Complesso Universitario di Monserrato - S.P. per Sestu Km 0.700,
	09042 Monserrato (Cagliari), Italy}

\author{N. Cargioli}
\email{nicola.cargioli@ca.infn.it}
\affiliation{Dipartimento di Fisica, Universit\`{a} degli Studi di Cagliari,
	Complesso Universitario di Monserrato - S.P. per Sestu Km 0.700,
	09042 Monserrato (Cagliari), Italy}
\affiliation{Istituto Nazionale di Fisica Nucleare (INFN), Sezione di Cagliari,
	Complesso Universitario di Monserrato - S.P. per Sestu Km 0.700,
	09042 Monserrato (Cagliari), Italy}

\author{F. Dordei}
\email{francesca.dordei@cern.ch}
\affiliation{Istituto Nazionale di Fisica Nucleare (INFN), Sezione di Cagliari,
	Complesso Universitario di Monserrato - S.P. per Sestu Km 0.700,
	09042 Monserrato (Cagliari), Italy}


\begin{abstract}
Elastic neutrino-electron scattering represents a powerful tool to investigate key neutrino properties. In view of the recent results released by the LUX-ZEPLIN collaboration, we provide a first determination of the limits achievable on the neutrino magnetic moment and neutrino millicharge, whose effect becomes non-negligible in some beyond the Standard Model theories. In this context, we evaluate and discuss the impact of different approximations to describe the neutrino interaction with atomic electrons. The new LUX-ZEPLIN data allows us to set a very competitive limit on the neutrino magnetic moment when compared to the other laboratory bounds, namely $\mu_{\nu}^{\rm{eff}} <  1.1 \times 10^{-11} \, \mu_{\text{B}}$ at 90$\%$ C.L., which improves by a factor of 2.5 the Borexino collaboration limit and represents the second best world limit after the recent XENONnT result.
Moreover, exploiting the so-called equivalent photon approximation, we obtain the most stringent limit on the neutrino millicharge, namely $|q_{\nu}^{\rm{eff}}| <  1.5 \times 10^{-13} e_0$ at 90$\%$ C.L., which represents a great improvement with respect to the previous laboratory bounds.
\end{abstract}

\maketitle  

\section{Introduction}
Recently, the LUX-ZEPLIN (LZ) Collaboration released the results~\cite{LZ} of the first search for so-called weakly interacting massive particles (WIMPs)~\cite{Lee:1977ua}, one of the most searched candidates to explain dark matter, which is predicted by a large number of theories beyond the Standard Model (SM)~\cite{Bertone:2016nfn,Billard:2021uyg,Akerib:2022ort}.  The LZ experiment is located at the Sanford Underground Research Facility in Lead, South Dakota. Its core is a dual-phase time projection chamber (TPC) filled with about 10 t of liquid xenon (LXe), of which 7 (5.5)~t of the active (fiducial) region. The possible interaction of a WIMP inside the detector produces two detectable signals if the nuclear recoil (NR) is above the $\sim$5 $\mathrm{keV}_{\mathrm{nr}}$ threshold, namely scintillation photons (S1) in the detector bulk and a secondary scintillation signal (S2) produced by the ionized electrons that drift thanks to an electric field to the gas pocket on top of the detector. Both signals are captured by 494 photomultiplier tubes located at the top and the bottom of the TPC. The results reported correspond to 60.3 live days and given that the data are consistent with a background-only hypothesis, permit setting the most stringent limits on the spin-independent and spin-dependent WIMP-nucleon scattering cross section for masses greater than 9~GeV/$c^2$~\cite{LZ}, as well as new competitive limits on the spin-dependent WIMP-proton cross section.\\

Among the different background components that characterize a direct dark matter experiment and that are kept into account in the data analysis, there is one due to elastic solar neutrino-electron scattering ($\nu$ES) inside the TPC. 
In the LZ analysis, the total number of such electron recoils (ERs) that is found after the combined fit of the background model plus a 30 GeV/$c^2$ WIMP signal is $27.2 \pm 1.6$~\cite{LZ} and represents about 10\% of the total background. Such a process is extremely sensitive to some neutrino electromagnetic properties beyond the SM (BSM), as the neutrino magnetic moment (MM) and the neutrino electric charge (EC), usually referred to as millicharge,  which can both significantly enhance the $\nu$ES contribution at low recoil energies~\cite{cadeddudresden,Coloma:2022avw,Borexino:2017fbd,Miranda:2020kwy,Grimus:2002vb,XENON:2020rca}. Thus, in this work, we revisit the fit to the LZ data allowing for a neutrino MM or a neutrino EC to set competitive limits on these quantities.

During the completion of this work, also the XENONnT collaboration reported its first result based on the analysis of low-energy ER data collected with a dual-phase TPC filled with 4.37~t of LXe fiducial mass and a total exposure of \mbox{1.16 t yr}~\cite{XENONCollaboration:2022kmb}. The experiment obtained the lowest ER background level among current dark matter detectors in its energy range of interest. No excess above the background is found, allowing the collaboration to rule out the well-known XENON-1T excess~\cite{XENON:2020rca}, most probably produced by an unaccounted tritium background. Moreover, they also reported a limit on the neutrino magnetic moment that will be compared to that obtained in this work.

\section{Theoretical framework}
\label{sec:cs}
Neutrino-electron elastic scattering is a source of background for direct searches of WIMPs. This background is in principle reducible,
but in practice hard to remove completely in experiments that use xenon due to the limited discrimination available between NRs and ERs. Luckily, in the SM its contribution to the total event rate at low recoil energies is rather precisely known and flat with respect to the recoil energy and thus it is usually subtracted in standard dark-matter analyses. However, in certain BSM scenarios, the $\nu$ES contribution could increase significantly, making it important to investigate this opportunity. Indeed, stronger constraints can be obtained on many neutrino electromagnetic properties~\cite{cadeddudresden,Coloma:2022avw,Borexino:2017fbd,Miranda:2020kwy,Grimus:2002vb}.\\

The SM $\nu$ES cross section per xenon atom is obtained multiplying the $\nu$ES cross section per electron with the effective electron charge of the target atom $Z_{\text{eff}}^{\mathrm{Xe}}(T_{e})$~\cite{cadeddudresden,Chen_2017,Kouzakov:2014lka}, and for each neutrino flavor $\nu_{\ell}$ ($\ell=e,\mu,\tau$) is given by
\begin{align}\nonumber
\dfrac{d\sigma_{\nu_{\ell}}}{d T_{\text{e}}}
(E,T_{\text{e}})
=
Z_{\text{eff}}^{\mathrm{Xe}}(T_{e})
\,
\dfrac{G_{\text{F}}^2 m_{e}}{2\pi}
[
\left( g_{V}^{\nu_{\ell}} + g_{A}^{\nu_{\ell}} \right)^2\\
+
\left( g_{V}^{\nu_{\ell}} - g_{A}^{\nu_{\ell}} \right)^2
\left( 1 - \dfrac{T_{e}}{E} \right)^2
-
\left( g_{V}^{\nu_{\ell}}\null^2 - g_{A}^{\nu_{\ell}}\null^2 \right)
\dfrac{m_{e} T_{e}}{E^2}
]
,
\label{ES-SM-flavour}
\end{align}
where $G_{\text{F}}$ is the Fermi constant, $E$ is the neutrino energy, $m_{e}$ is the electron mass, $T_e$ is the electron recoil energy, and the neutrino-flavor dependent electron couplings at tree level are
\begin{align}
\null & \null
g_{V}^{\nu_{e}}
=
2\sin^{2} \theta_{W} + 1/2
,
\quad
\null && \null
g_{A}^{\nu_{e}}
=
1/2
,
\label{gnue}
\\
\null & \null
g_{V}^{\nu_{\mu,\tau}}
=
2\sin^{2} \theta_{W} - 1/2
,
\quad
\null && \null
g_{A}^{\nu_{\mu,\tau}}
=
- 1/2
.
\label{gnumu}
\end{align}
They correspond to $g_{V}^{\nu_{e}}=0.9521$, $g_{A}^{\nu_{e}}=0.4938$, $g_{V}^{\nu_{\mu}}=-0.0397$, $g_{A}^{\nu_{\mu,\tau}}=-0.5062$, and $g_{V}^{\nu_{\tau}}=-0.0353$ when taking into account radiative corrections (see Appendix~\ref{app:rad} for further information).
Here, $\theta_{W}$ is the weak mixing angle, also known as the Weinberg angle, whose value at zero momentum transfer is $\sin^{2} \theta_{W}=0.23857$~\cite{ParticleDataGroup:2020ssz} in the $\overline{\mathrm{MS}}$ scheme.
The $Z_{\text{eff}}^{\mathrm{Xe}}(T_{e})$ term~\cite{Mikaelyan:2002nv,Fayans:2000ns} quantifies the number of electrons that can be ionized by a certain energy deposit $T_e$ and is needed to correct the cross section derived under the free electron approximation (FEA) hypothesis. This is especially important for Xe, where one expects a rather big effect from atomic binding~\cite{Chen_2017}.
It has been obtained by using the edge energies extracted from photoabsorption data~\cite{Chen_2017,HENKE1993181} (see Appendix~\ref{app:zeff} for further information). 
An alternative method implies the usage of the so-called relativistic random-phase approximation (RRPA) theory~\cite{Chen_2017,PhysRevA.25.634,PhysRevA.26.734,Chen:2013lba}. With respect to the FEA corrected with the stepping function $Z_{\text{eff}}^{\mathrm{Xe}}(T_{e})$, RRPA provides an {\textit{ab initio}} approach able to give an improved description
of the atomic many-body effects. In the case of neutrino SM interactions or with additional neutrino MMs, it slightly reduces the $\nu$ES number of events by an almost constant value as a function of the recoil energy.
On the other hand, in the case of neutrino ECs, the low-energy ER spectrum is highly enhanced when using the RRPA formalism with respect to the corrected FEA approach. In this particular case, it is also possible to use the equivalent photon approximation (EPA), which relates the ionization cross section to the photo-absorption one, reproducing closely the RRPA cross section for a millicharged neutrino~\cite{Chen:2014ypv,Hsieh:2019hug}.\\

The total SM differential cross section includes the contribution from all neutrino flavors keeping into account the oscillation probability in the three-neutrino oscillation scheme and it is
\begin{equation}
\dfrac{d\sigma_{\nu}}{d T_{\text{e}}}
(E,T_{\text{e}})=
P_{ee}\dfrac{d\sigma_{\nu_{e}}}{d T_{\text{e}}}
+\sum_{f=\mu,\tau}P_{ef}\dfrac{d\sigma_{\nu_{f}}}{d T_{\text{e}}}
,
\label{ES-SM}
\end{equation}
where $P_{ee}=\sin^4\theta_{13}+\cos^4\theta_{13}P^{2\nu}$~\cite{Borexino:2017fbd} is the average survival probability for solar neutrinos reaching the detector when considering the dominant $pp$ and $^7$Be fluxes and $P^{2\nu}\simeq  0.55$~\cite{Chen_2017,ParticleDataGroup:2020ssz} is the $\nu_e$ survival probability in the two-neutrino oscillation scheme. Here, $P_{e\mu}=(1-P_{ee})\cos^2\theta_{23}$ and $P_{e\tau}=(1-P_{ee})\sin^2\theta_{23}$ are the transition probabilities. The values of the corresponding mixing angles $\theta_{13}$ and $\theta_{23}$ were taken from Ref.~\cite{ParticleDataGroup:2020ssz}.\\


\section{Neutrino magnetic moment}
\label{sec:magnetic}
In the SM, neutrinos are considered massless, and therefore neutrino MMs are vanishing. Nevertheless, from the fact that neutrino oscillates, we know that the
SM must be extended to give masses to the neutrinos. In the minimal extension of the SM in which neutrinos acquire Dirac masses through the introduction of right-handed neutrinos, the neutrino MM
is given by~\cite{Giunti:2014ixa,Giunti:2015gga,PhysRevLett.45.963,Schechter:1981hw,Kayser:1982br,Nieves:1981zt,Pal:1981rm,Shrock:1982sc}
\begin{equation}
    \mu_\nu = \frac{3e_0G_F}{8\sqrt{2}\pi^2} m_\nu \simeq 3.2 \times 10^{-19} \left( \frac{m_\nu}{\mathrm{eV}} \right)\mu_B,
\end{equation}
where $\mu_B$ is the Bohr magneton, $m_\nu$ is the neutrino mass and $e_0$ is the electric charge. 
Taking into account the current upper limit on the neutrino mass~\cite{ParticleDataGroup:2020ssz}, this value is less than $\mu_\nu \sim 10^{-18} \mu_B$, which is too small to be observed experimentally.
Nevertheless, given that in some BSM scenarios the neutrino MM is predicted to be larger~\cite{Giunti:2015gga}, a positive observation would represent a clear
signal of physics beyond the minimally extended SM. For this reason, neutrino MM is the most investigated neutrino electromagnetic
property, both theoretically and experimentally.

An enhanced MM would increase the
neutrino scattering cross sections at low energies on
both electrons and nuclei, and thus could be observable
by low-threshold detectors, such as the liquid xenon dark matter detectors, as discussed in Refs. \cite{Huang_2019,Babu:2020ivd,PhysRevD.106.015002,Li:2022bqr}. 
By considering the enhancement due to $\nu$ES, the differential $\nu$ES cross section that takes into account the contribution of the neutrino MM is given by adding to the SM cross section in Eq.~(\ref{ES-SM-flavour}) the MM contribution, namely
\begin{equation}
\dfrac{d\sigma_{\nu_{\ell}}^{\text{MM}}}{d T_\mathrm{e}}
(E,T_\mathrm{e})=
Z_{\text{eff}}^{\mathrm{Xe}}(T_{\text{e}}) \dfrac{ \pi \alpha^2 }{ m_e^2 }
\left( \dfrac{1}{T_\mathrm{e}} - \dfrac{1}{E} \right)
\left| \dfrac{\mu_{\nu_{\ell}}}{\mu_{\text{B}}} \right|^2,
\label{es-mag}
\end{equation}
where $\mu_{\nu_{\ell}}$ is the effective MM of the flavor neutrino $\nu_{\ell}$
in elastic scattering (see Ref.~\cite{Giunti:2014ixa}).\\


\section{Neutrino millicharge}
It is usually believed that neutrinos are neutral particles. However, in some BSM theories they can acquire a small electric
charge (see Ref.~\cite{Giunti:2014ixa} and references therein).
Within the FEA approach corrected by the stepping function, the millicharged neutrino contribution to the differential ES cross section can be obtained by modifying the neutrino vector coupling $g_V^{\nu_\ell}$ in Eq.~(\ref{ES-SM-flavour})  through  
\begin{equation}
    g_V^{\nu_\ell}\rightarrow g_V^{\nu_\ell}+\frac{2\sqrt{2}\pi\alpha}{G_F q^2}q_{\nu_\ell},
\end{equation}
where $q_{\nu_\ell}$ is the EC associated to the flavor $\ell$ and $q^2=-2m_eT_e$ is the momentum transfer in the interaction. Let us note that in this case, the neutrino EC can interfere with the SM coupling so that the sign of the electric charge is important, while the MM correction is independent of the sign.
Given that for low ER energies the momentum transfer is small, the analysis of the LZ data is expected to be particularly promising for millicharged neutrino searches. 
It is worth mentioning that, although the neutrino MM cross section within the corrected FEA framework is known to be in good agreement with that of {\textit{ab initio}} theories even for sub-keV ERs, in the same regime the RRPA cross section for a neutrino EC is more than one order of magnitude bigger than that obtained with the corrected FEA~\cite{Chen:2014ypv,Hsieh:2019hug}. In this regard, we can consider the neutrino EC limit obtained within the FEA formalism as a conservative one. 
Given that it is well known that the EPA scheme reproduces well the RRPA cross section for a millicharged neutrino~\cite{Chen:2014ypv,Hsieh:2019hug}, we exploit the EPA formalism in order to go beyond the FEA approach and better describe the interaction. This improved approach should lead to tighter constraints on the neutrino millicharge. In particular, the EPA cross section for a millicharged ultrarelativistic particle reads~\cite{Chen:2014ypv,Hsieh:2019hug}

\begin{equation}
    \dfrac{d\sigma_{\nu_\ell}}{d T_\text{e}}\Big\vert_{\rm{EPA}}^{\rm{EC}}=\frac{2\alpha}{\pi}\frac{\sigma_\gamma(T_e)}{T_e}\log\left[\frac{E_\nu}{m_\nu}\right]q_{\nu_\ell}^2
    \label{eq:EPA},
\end{equation}

where $m_\nu$ is the neutrino mass, and $\sigma_\gamma(T_e)$ is the photoelectric
cross section by a real photon, which can be extracted from Ref.~\cite{HENKE1993181} for Xe.
By looking at Eq.~(\ref{eq:EPA}) it can be seen that the cross section in the EPA approximation is independent of the sign of the electric charge, differently from the case of the FEA approximation.
We should underline that, although the EPA approach describes very well the cross section for ER energies below a few keVs, it is known to underestimate the scattering cross section for larger energies where the FEA formalism works better. For this reason, we will rely on the EPA scheme only when its cross section is larger than that of the corrected FEA, following the same procedure adopted in Ref.~\cite{TEXONO:2018nir}. In the following, for simplicity, we will refer to this strategy as EPA.
\\


\section{Data analysis strategy}
\label{sec:method}
For the analysis of the LZ dataset,  we obtained information on all the quantities used from Ref.~\cite{LZ} and the accompanying data release and supplemental material unless noted otherwise.\\

The total differential neutrino flux, $d N_{\nu,j}/d E$, is given by the sum of all the different solar neutrino components $j$ as from Refs.~\cite{ParticleDataGroup:2020ssz,Vitagliano:2019yzm}, of which the most relevant for the sensitivity range of LZ
are the continuous $pp$ flux and the monochromatic $^{7}$Be 861~keV line, even though there are many additional contributions from other mechanisms that are included in the analysis. \\

\begin{figure}
    \centering
    	\includegraphics[width=\columnwidth]{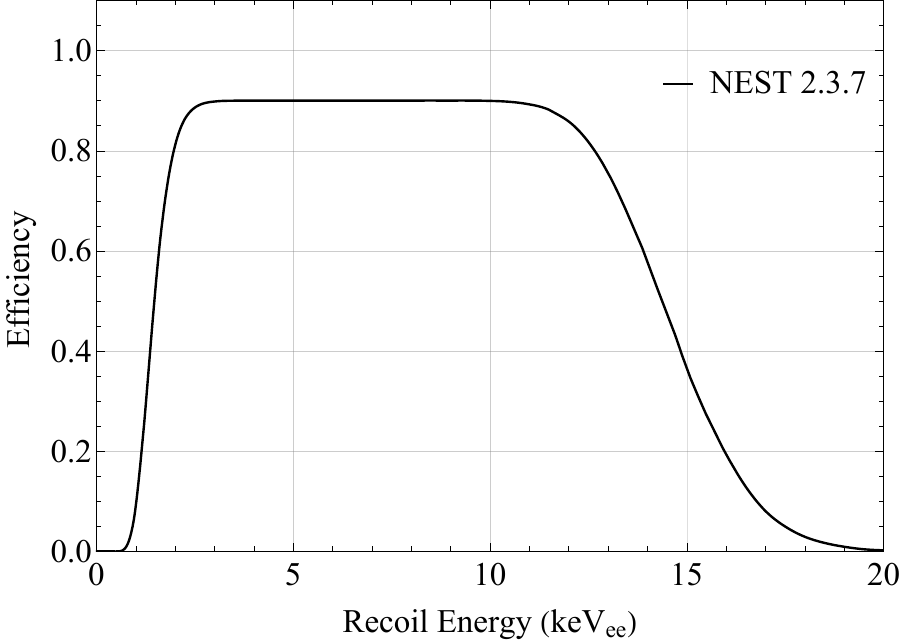}
    \caption{LZ signal efficiency as a function of the ER energy $T_e$, obtained from the \textsc{NEST} 2.3.7 software using the details provided by the LZ collaboration.}
    \label{fig:Eff_ER}
\end{figure}

In each ER energy bin $i$, the theoretical $\nu$ES event number $N_i^{\rm \nu ES}$  is given by
\begin{align}\label{N_es}\nonumber
N_i^{\rm \nu ES}
&=
N(\mathrm{Xe})
\int_{T_{\mathrm{e}}^{i}}^{T_{\mathrm{e}}^{i+1}}
\hspace{-0.1cm}
d T_{\mathrm{e}}\,
A(T_{\mathrm{e}})\\
&\int_{E_{\text{min}}(T_{\text{e}})}^{E_{\text{max}}}
\hspace{-0.3cm}
d E
\sum_{j}
\frac{d N_{\nu,j}}{d E}(E)
\frac{d \sigma_{\nu}}{d T_{\mathrm{e}}}(E, T_{\mathrm{e}})
,
\end{align}
where $N(\mathrm{Xe})$ is the number of xenon targets contained in the detector, $T_{\text{e}}$ is the ER kinetic energy,
$A(T_{\text{e}})$ is the energy-dependent detector efficiency,
$E_{\text{min}}(T_{\text{e}}) = (T_\text{e}+\sqrt{T_\text{e}^{2} + 2m_e T_\text{e}})/2$, and $E_{\text{max}} \sim 2$~MeV.
The number of target xenon atoms in the detector is given by
$N(\mathrm{Xe}) = N_{\mathrm{A}} M_{\mathrm{det}} / M_{\mathrm{\mathrm{Xe}}}$, 
where $N_{\mathrm{A}}$ is the Avogadro number, 
$M_{\mathrm{det}}=5.5~\mathrm{t}$ is the detector fiducial mass and
$M_{\mathrm{\mathrm{Xe}}}$ is the average xenon molar mass.\\
While the LZ collaboration provided the detector efficiency as a function of the NR energy $T_{\mathrm{nr}}$, the energy observed in the detector is the ER energy $T_{e}$. For this reason, we derived the detector efficiency as a function of $T_{e}$ using the \textsc{NEST~\cite{Szydagis_2011}} 2.3.7 software, following the information provided by the LZ collaboration. The efficiency obtained and used in our analysis is shown in Fig.~\ref{fig:Eff_ER}.
It is worth mentioning that in a preliminary version of this work~\cite{AtzoriCorona:2022jeb} we retrieved the ER efficiency curve by converting the NR one through the Lindhard quenching factor~\cite{osti_4701226}. However, this procedure, which was also employed in Ref.~\cite{Maity:2022exk}, neglects the different contributions from ionization and scintillation channels in a dual-phase TPC. Hence, it is not correct and it leads to an incorrect ER efficiency. In particular, the latter procedure overestimates the ER efficiency at low energies, lowering the threshold and hence, leading erroneously to a much stronger sensitivity to the new-physics scenarios considered.\\

Besides the solar $\nu$ES, the background components that survive the selection in the region of interest  come from different sources, the dominant one being the ERs from radioactive decay of impurities dispersed in the xenon, commonly referred to as $\beta$ background. Together with a small ($<1\%$) fraction due to ER from $\gamma$ rays originating in the detector components and cavern walls, this background represents about 79\% of the total one. Other background sources include the naturally occurring isotopes of xenon, which also contribute to ER events, as well as isotopes that are activated cosmogenically, such as $^{127}$Xe and $^{37}$Ar. Moreover, the NR background has contributions from radiogenic neutrons and coherent elastic neutrino-nucleus scattering (CE$\nu$NS) from $^8$B solar neutrinos. Finally, there is a small component of accidental backgrounds that is also kept into account. Overall, the LZ collaboration reports a background of $333 \pm 17$ events, of which $27.2 \pm 1.6$ are due to solar $\nu$ES, see Table I in Ref.~\cite{LZ}.\\

We performed the analysis of the LZ data using a Poissonian least-squares function~\cite{Baker:1983tu,ParticleDataGroup:2020ssz}, given that in some energy bins the number of events is small, namely
\begin{align}\nonumber
\label{eq:chi2xe}
    \chi^2
    & =
    	\null  
    2\sum_{i=1}^{51}
    [
    (1+\alpha) N_i^{\rm bkg}
    + (1+\beta) N_i^{\rm \nu ES}
    - N_i^{\rm exp}\\
    &+ N_i^{\rm exp} \ln\left(\frac{N_i^{\rm exp}}{(1+\alpha) N_i^{\rm bkg}
    + (1+\beta) N_i^{\rm \nu ES}}\right)
    ]\\\nonumber
	&+\Big(\dfrac{\alpha}{\sigma_\alpha}\Big)^2
    +\Big(\dfrac{\beta}{\sigma_\beta}\Big)^2,
\end{align}
where $N_i^{\rm bkg}$ is the number of residual background events found in the $i$th bin fit by the LZ collaboration minus that due to solar $\nu$ES (both extracted from Fig.~6 of Ref.~\cite{LZ}), $N_i^{\rm \nu ES}$ is the prediction in the $i$th bin for the $\nu$ES signal, and $N_i^{\rm exp}$ is the experimental number of events in the $i$th bin, also extracted from Fig.~6 of Ref.~\cite{LZ}. The nuisance parameter $\alpha$ takes into account the uncertainty on the neutrino background (with $\sigma_\alpha=5.1\%$)\footnote{We note that this procedure ignores the fact that the different background contributions have a different relative uncertainty. However, given that the total background is dominated by the $\beta$ decays this approximation is valid.}, while $\beta$ keeps into account the uncertainty on the neutrino flux (with $\sigma_\beta=7\%$)\footnote{The flux uncertainty is about 7\% for $^7$Be and 0.6\% for $pp$~\cite{Serenelli:2016dgz}, we conservatively use the first one for both fluxes.}. By using this procedure we ignore that a possible nonzero neutrino MM should also increase the CE$\nu$NS contribution from $^8$B solar neutrinos. However, given that the latter contribution is only $0.15 \pm 0.01$, we verified that we can safely neglect it. For the future, we note that a lower experimental energy threshold would increase the CE$\nu$NS contribution, thus contributing to further strengthening the MM and EC limits.\\

We highlight that, differently from all the other background sources, the number of $^{37}$Ar events is not well constrained theoretically. It
is estimated by calculating the exposure of Xe to
cosmic rays before it was brought underground, then correcting for the decay time before the search~\cite{LUX-ZEPLIN:2022sad}. A flat
constraint of 0 to three times (i.e., 288) the estimate of 96 events is imposed because of large uncertainties in the prediction. The fit to the data using this prior finds $52.5^{+9.6}_{-8.9}$ events. In order to keep into account this large uncertainty, we perform a second analysis in which we separate the $^{37}$Ar contribution from the total background such that the least-squares function becomes
\begin{align}\nonumber
\label{eq:chi2xear37}
    \chi^2_{{^{37}}\mathrm{Ar}}
    & =
    2\sum_{i=1}^{51}[
    \alpha N_i^{\rm bkg}
    + \beta N_i^{\rm \nu ES}
    + \delta N_i^{\rm {^{37}}\mathrm{Ar}}
    - N_i^{\rm exp} \\\nonumber
    & + N_i^{\rm exp} \ln\left(\frac{N_i^{\rm exp}}{\alpha N_i^{\rm bkg}
    + \beta N_i^{\rm \nu ES} + \delta N_i^{\rm {^{37}}\mathrm{Ar}}}\right)
    ]\\
	&+\Big(\dfrac{\alpha-1}{\sigma_\alpha}\Big)^2
    +\Big(\dfrac{\beta-1}{\sigma_\beta}\Big)^2
    +\Big(\dfrac{\delta-1}{\sigma_\delta}\Big)^2,
\end{align}

where 
$N_i^{\rm bkg}$ is the number of residual background events minus those due to $\nu$ES and $^{37}$Ar as found in the $i$th electron recoil energy bin fit by the LZ collaboration, and $N_i^{\rm {^{37}}\mathrm{Ar}}$ is the number of $^{37}$Ar background events found in the $i$th bin fit by the LZ collaboration, scaled such that the integral is equal to 96 events, as estimated in Ref.~\cite{LZ}. We leave the latter free to vary in the fit with a Gaussian constraint given by the nuisance parameter $\delta$, which takes into account the uncertainty on the $^{37}$Ar background, with $\sigma_\delta=100\%$. In this case, we set $\sigma_\alpha=13\%$, which is the uncertainty on the expected number of background events provided in Ref.~\cite{LZ} when not considering the $^{37}$Ar contribution.\\

\begin{figure}
    \centering
    	\includegraphics[width=\columnwidth]{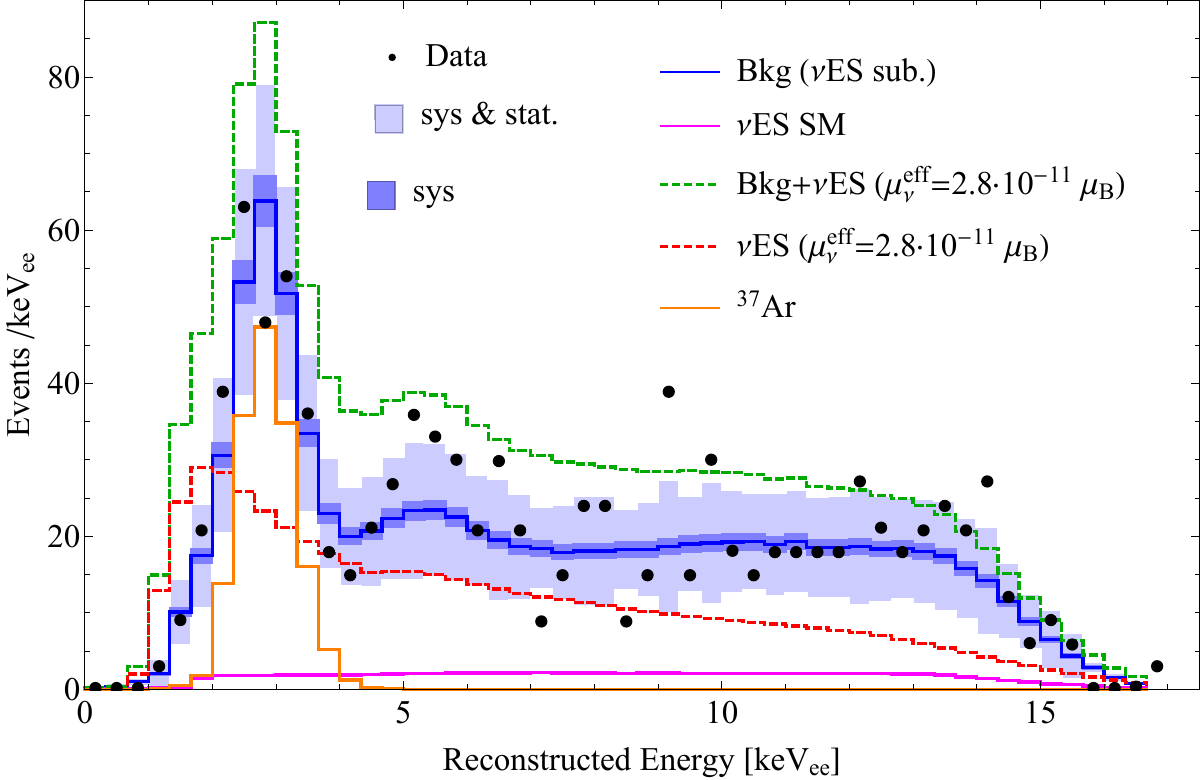}
    \caption{
    LZ energy spectrum (black points) as extracted from Fig.~6 of Ref.~\cite{LZ} with superimposed the sum of all background contributions minus the $\nu$ES contribution (blue solid), the $^{37}$Ar contribution (orange), the $\nu$ES SM prediction (purple), and for illustration purposes the $\nu$ES with $\mu_{\nu}^{\rm eff}=2.8\times10^{-11}~\mu_{\text{B}}$, that corresponds to the 90\% C.L. limit from BOREXINO~\cite{Borexino:2017fbd}, with (green dashed) and without (red dashed) the $\nu$ES subtracted background. The dark blue and the light blue bands represent the systematic and systematic plus statistical uncertainties, respectively, used in this analysis.
    }
    \label{fig:spe_LZ}
\end{figure}

In Fig.~\ref{fig:spe_LZ} we show an example of the $\nu$ES prediction in presence of a possible neutrino MM for the LZ spectrum compared with the data, the SM $\nu$ES prediction and the other background components, considering e.g. $\mu_{\nu}^{\rm eff}=2.8\times10^{-11}~\mu_{\text{B}}$, which corresponds to the previous best limit at 90 $\%$ confidence level (C.L.) on the neutrino MM from Borexino~\cite{Borexino:2017fbd}.\\

\section{Results}
\label{sec:result}
Since neutrinos are a mixture of mass eigenstates due to the phenomenon of oscillations, the
MM measured for solar $\nu$ES is an effective value given by
\begin{equation}
\mu^{2,\rm{eff}}_\nu = \sum_j | \sum_k \mu_{jk}A_k(E_\nu,L)|^2,
\end{equation}
where $\mu_{jk}$ is an element of the neutrino electromagnetic moments matrix and $A_k(E_\nu,L)$ is the amplitude of
the $k$-mass state at the point of scattering~\cite{Borexino:2017fbd}. For the
Majorana neutrino, only the transition moments are nonzero, while the diagonal elements of the matrix are equal to zero due to CPT conservation. For the Dirac neutrino,
all matrix elements may have nonzero values~\cite{Kouzakov:2017hbc}. \\
Similarly, it is possible to define also an effective neutrino millicharge parameter $q_{\nu}^{\rm eff}$ as a combination of the three flavor components. \\

\begin{figure}
    \centering
    	\includegraphics[width=\columnwidth]{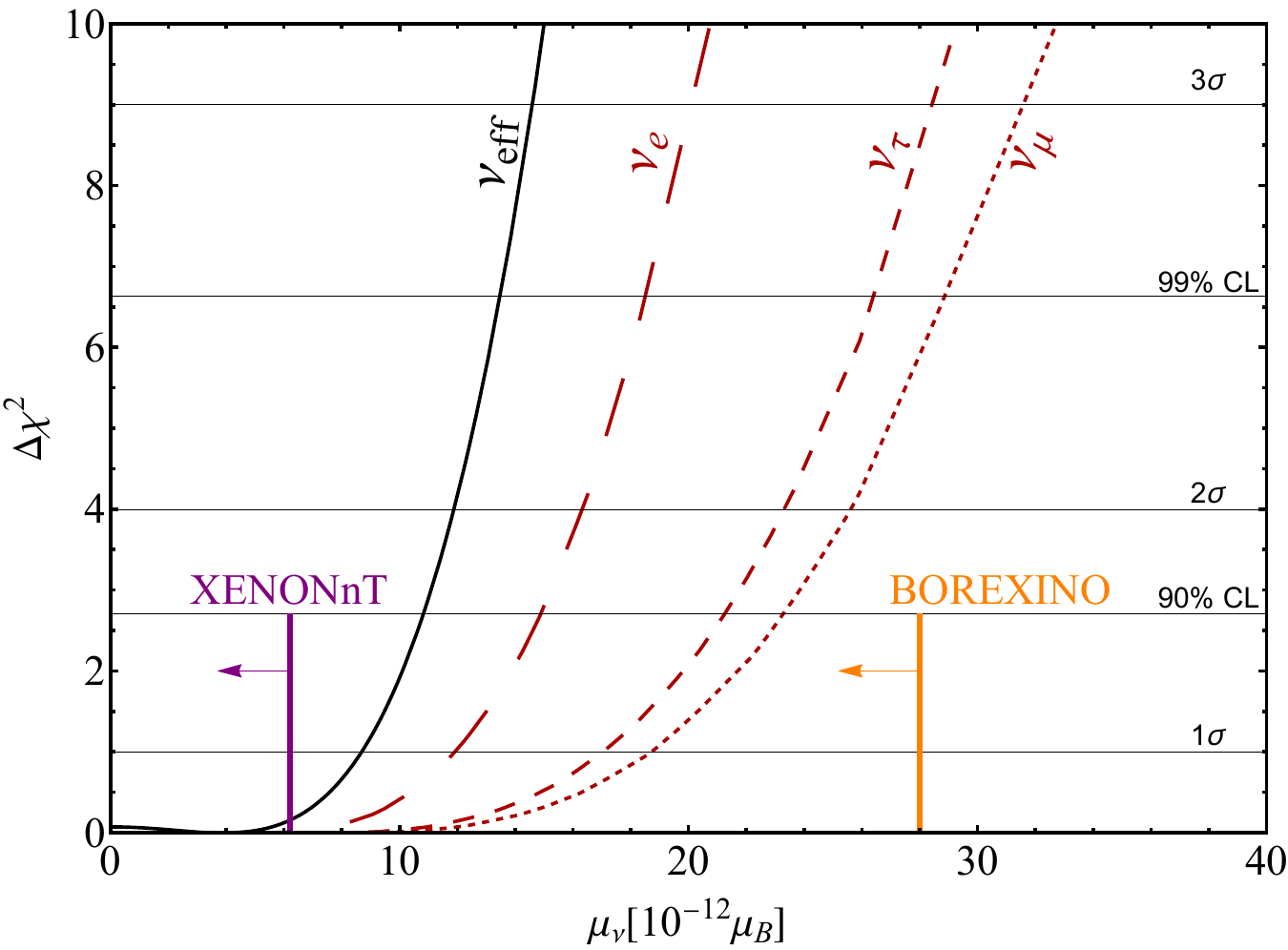}
    \caption{
    Marginal $\Delta\chi^2$s for $\mu_{\nu}^{\rm{eff}}$ obtained from the analysis of the LZ data with the $\chi^2$ in Eq.~(\ref{eq:chi2xe}) (black solid line) and the marginalized flavor components (dashed red lines). The solid purple (orange) line shows the 90\% C.L. upper bound on the effective neutrino MM obtained in the XENONnT~\cite{XENONCollaboration:2022kmb} (BOREXINO~\cite{Borexino:2017fbd}) experiment.}
\label{fig:magLZ}
\end{figure}

In Fig.~\ref{fig:magLZ} we show the marginal $\Delta\chi^2$s at different confidence levels, obtained using the $\chi^2$ in Eq.~(\ref{eq:chi2xe}), for both the effective MM and the marginalization over the three flavor components. 
\begin{table}
{
\input{tab_limits.tex}
}
\caption{Limits on the neutrino magnetic moment and neutrino millicharge at 90$\%$ C.L. obtained with a $\chi^2$ analysis as defined in Eq.~(\ref{eq:chi2xe}). For the neutrino millicharge, the limits are reported for both the FEA and the EPA formalism.}  \label{tab:tablimits}
\end{table}
The numerical values of the limits derived considering the three different flavors are reported in Table~\ref{tab:tablimits}.
At $90\%$ C.L., the bound on the effective neutrino MM obtained in this work is
\begin{eqnarray}
    \mu_{\nu}^{\rm{eff}}& < & 1.1 \times 10^{-11} \, \mu_{\text{B}},
\end{eqnarray}
with the minimum of the chi-square being $\chi^2_{\rm min}=100.0$, which corresponds to an integrated number of $\sim$50 $\nu$ES events.
It can be compared with the limit recently reported by the XENONnT collaboration corresponding to $\mu_{\nu}^{\rm{eff}} <  6.4 \times 10^{-12} \, \mu_{\text{B}}$~\cite{XENONCollaboration:2022kmb}, which is about a factor of 2 more stringent due to their lower background with respect to LZ. Further neutrino MM analyses exploiting XENONnT data can be found in Refs.~\cite{A:2022acy,Khan:2022bel}. These LZ and XENONnT limits, both obtained using a LXe double-phase TPC technology originally designed to search for dark matter and a similar analysis approach, are significantly tighter than the previous laboratory bounds, highlighting the potentiality that such a technique can offer thanks to the low energy threshold and low level of background achieved. Indeed, they can be compared to the limit obtained by the Super-Kamiokande collaboration of $3.6\times 10^{-10} \mu_B$ (90\% C.L.), derived by
fitting day/night solar neutrino spectra above 5~MeV.
With additional information from other solar neutrino
and KamLAND experiments a limit of $1.1\times 10^{-10} \mu_B$ (90\%
C.L.) was obtained~\cite{Super-Kamiokande:2004wqk}. The Borexino collaboration reported the previous best current limit on the effective MM by laboratory experiments of $2.8\times 10^{-11} \mu_B$ (90\% C.L.) using the ER spectrum from solar neutrinos~\cite{Borexino:2017fbd}.
The best MM limit from reactor antineutrinos is $2.9\times 10^{-11} \mu_B$ (90\% C.L.)~\cite{Gemma}. Finally, the analysis of the CE$\nu$NS data from Dresden-II and COHERENT collaborations permits to set limits on $|\mu_{\nu_{e}}| <  2.13 \times 10^{-10} \, \mu_{\text{B}} $ and $|\mu_{\nu_{\mu}}|  <  18 \times 10^{-10} \, \mu_{\text{B}}$~\cite{cadeddudresden}, also exploiting $\nu$ES.
When considering nonlaboratory experiments, the most
stringent limits on the neutrino MM of up to $\sim 10^{-12} \mu_B$ come from astrophysical observations~\cite{ARCEODIAZ20151,Diaz:2019kim,Corsico:2014mpa}, which however are rather indirect. A complete historical record of limits on the
neutrino MM can be found in Ref.~\cite{ParticleDataGroup:2020ssz} and a large collection of existing bounds is summarized in Fig.~\ref{fig:GS}(a). It is possible to see that in our analysis of the LZ data we significantly improve the limits on the electron, muon and tau neutrino MM compared to the other laboratory bounds.

\begin{figure}[t]
\includegraphics[width=0.49\textwidth]{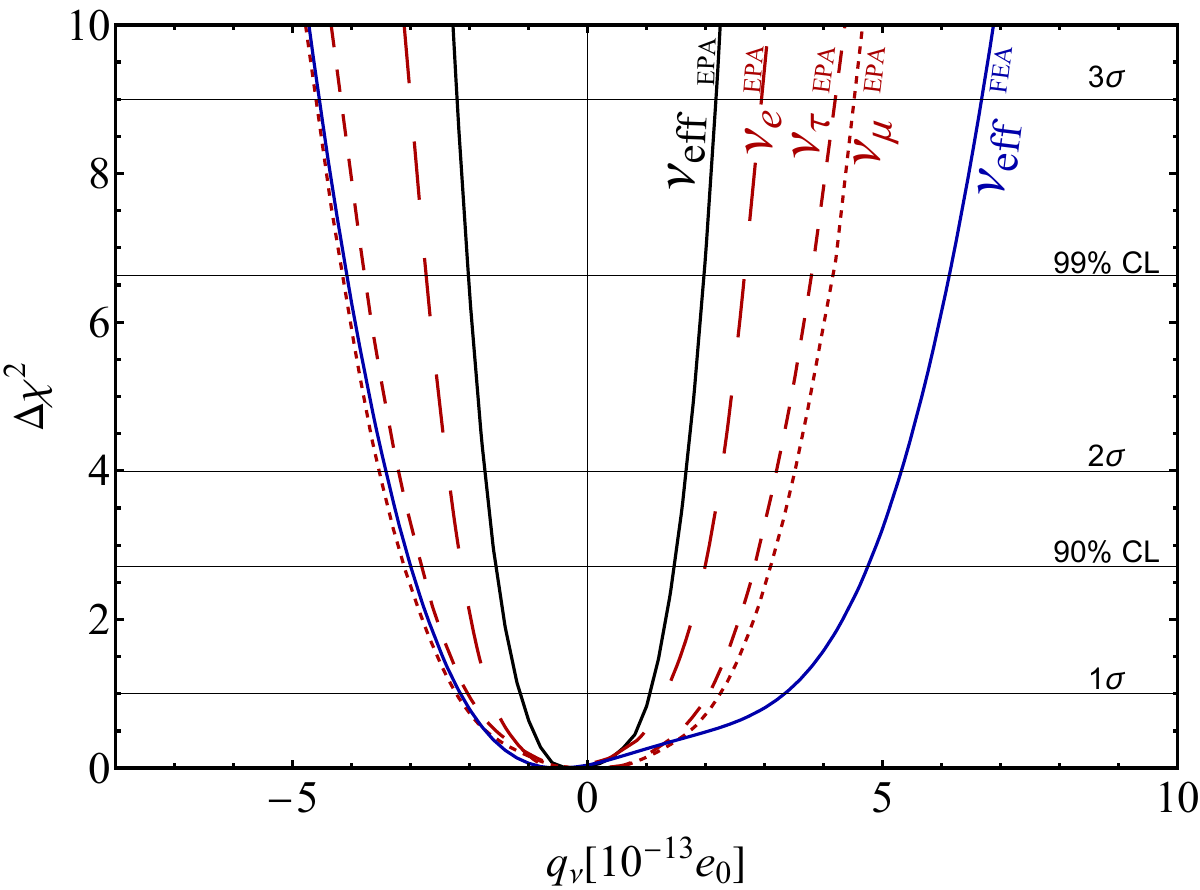}
\caption{$\Delta\chi^2$ profiles of the effective (solid black) and flavor dependent (dashed red) neutrino millicharge obtained adopting the EPA formalism. As a comparison, the curve for the effective neutrino millicharge under the FEA approximation is also shown (solid blue).}
\label{fig:Millicharge}
\end{figure}

\begin{figure*}
    \centering
    	\includegraphics[width=\textwidth]{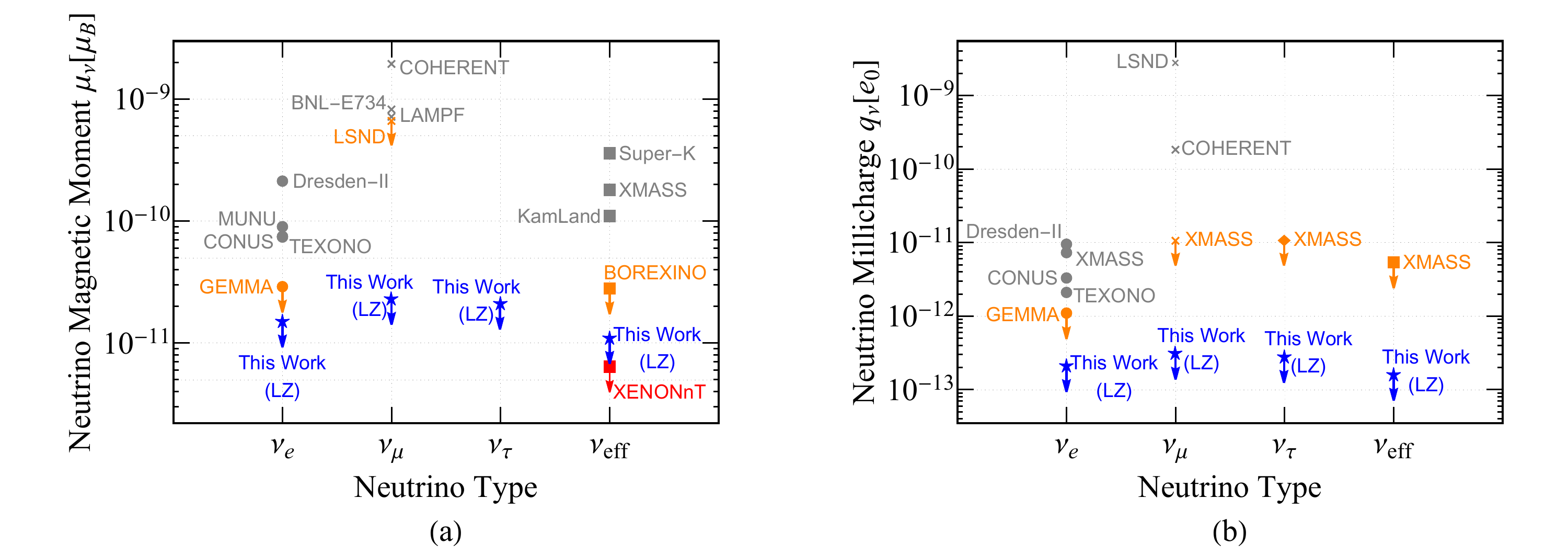}
    \caption{
    Summary of existing limits at 90$\%$ C.L. on the neutrino magnetic moment (a) and the neutrino millicharge (b) coming from a variety of experiments~\cite{Beda:2012zz,TEXONO:2006xds,Borexino:2017fbd,Super-Kamiokande:2004wqk,cadeddudresden,CONUS:2022qbb,MUNU:2005xnz,Allen:1992qe,Ahrens:1990fp,LSND:2001akn,Giunti:2014ixa,XMASS:2020zke,ParticleDataGroup:2020ssz}.The limits are divided in flavor components $\mu_{\nu_e}$ ($q_{\nu_e}$) (dots), $\mu_{\nu_\mu}$ ($q_{\nu_\mu}$) (crosses), and $\mu_{\nu_\tau}$ ($q_{\nu_\tau}$) (diamonds) and also the ones on the effective magnetic moment $\mu_{\nu}^{\rm eff}$ ($q_{\nu}^{\rm eff}$) (squares) are shown. In orange, we highlighted the best limits before the LZ data release and in red the XENONnT limit on the MM~\cite{XENONCollaboration:2022kmb}. The results derived in this work for the effective parameter as well as divided in flavors are shown by the blue stars.}
    \label{fig:GS}
\end{figure*}

We checked the impact on the neutrino MM limits of introducing the detector energy resolution in Eq.~(\ref{N_es}), which is measured to be very precise by the LZ collaboration. For this check, the theoretical spectra were smeared using a Gaussian distribution with an energy-dependent width, which has been determined using an empirical fit of mono-energetic peaks~\cite{LZ:2019sgr}. In particular, for the latter we employed the value reported in Ref.~\cite{resosupp}, namely $\sigma (T_e) = K/\sqrt{T_e}$, with $K=0.323\pm0.001$. Thanks to the excellent energy resolution achieved by LZ, we verified that its inclusion does not significantly modify the limit obtained.
Finally, we investigated the possibility of leaving the $^{37}$Ar component free to vary in the fit using a prior similar to that implemented by the LZ collaboration, as defined in Eq.~(\ref{eq:chi2xear37}). Interestingly, the fit retrieves a number of $^{37}$Ar events similar to that found by LZ, namely $\sim 48$ with $\chi^2_{\rm min}=99.6$. Thus, also in this case, the limits do not substantially change and for reference the bound on the effective neutrino MM at 90$\%$ C.L. becomes
    $\mu_{\nu}^{\rm{eff}} (^{37}\mathrm{Ar}) <  1.2 \times 10^{-11} \, \mu_{\text{B}}$.\\ \\
As stated in the introduction, the LZ dataset is also very sensitive to a possible neutrino millicharge. In Fig.~\ref{fig:Millicharge} we present the limits on the neutrino EC obtained in this work within the FEA and EPA formalisms, using the $\chi^2$ in Eq.~(\ref{eq:chi2xe}). We note that the EPA cross section depends on the neutrino mass, as it can be seen in Eq.~(\ref{eq:EPA}), which is not yet precisely measured. We used a conservative value of $m_\nu=1$~eV, which is close to the current laboratory upper bounds on the neutrino mass~\cite{ParticleDataGroup:2020ssz}. On the other hand, we verified that the limit is not significantly modified even when considering smaller values for $m_\nu$. The 90$\%$ C.L. bounds on the effective millicharge are
\begin{align}
    &{\rm{FEA:}}\,\,\,\,-3.0 <  q_{\nu}^{\rm{eff}}\;[10^{-13}\;e_0] <  4.7,\\
    &
    {\rm{EPA:}}\,\,\,\,-1.5 <  q_{\nu}^{\rm{eff}}\;[10^{-13}\;e_0] <  1.5,
\end{align}
the minimum of the chi-square being $\chi^2_{\rm min}=100.0$ in both cases. The values for the flavor-dependent neutrino millicharges are summarized in Table~\ref{tab:tablimits} both for the FEA and EPA analyses.
It is clear that the limits obtained with the more realistic EPA formalism are much stronger than those obtained within FEA and hence, for simplicity, in Fig.~\ref{fig:Millicharge} we showed only the effective EC limit for FEA. 
We note also that the limits obtained in this work with FEA are comparable with those reported in Ref.~\cite{A:2022acy}, which exploits the ER energy efficiency derived in this work for the LZ analysis, and are less stringent than those obtained with XENONnT~\cite{A:2022acy,Khan:2022bel}.
On the other hand, as expected, the limits obtained in this work adopting EPA when analyzing the LZ data are even stronger than the XENONnT limits obtained in Refs.~\cite{A:2022acy,Khan:2022bel} that were determined using FEA.

In Fig.~\ref{fig:GS}(b) a collection of existing bounds coming from different experiments is shown. It can be seen that the limits derived in this work using the LZ data and the more realistic EPA formalism significantly improve the previous best laboratory limits, that for the electron neutrino electric charge was obtained in Ref.~\cite{Chen:2014dsa} by combining TEXONO~\cite{TEXONO:2006xds} and GEMMA~\cite{Beda:2012zz} data, finding $|q_{\nu_e}|< 1.0\times10^{-12}\;e_0$.
We expect, however, that adopting the EPA or the RRPA formalism to analyse the XENONnT data would allow us to further constrain the limit on this fundamental quantity. This investigation will be carried out in a future work.

For completeness, also in this case we investigated the impact of repeating the analysis leaving the $^{37}$Ar component free to vary, similarly to what was done for the neutrino MM limits. In this case, the bounds on the effective neutrino millicharge become
\begin{align}
    &{\rm{FEA:}}\,\,\,\,-3.3 <  q_{\nu}^{\rm{eff}}(^{37}\mathrm{Ar})\;[10^{-13}\;e_0] <  5.0,\\
    &
    {\rm{EPA:}}\,\,\,\,-1.6 <  q_{\nu}^{\rm{eff}}(^{37}\mathrm{Ar})\;[10^{-13}\;e_0] <  1.5,
\end{align}
with the minimum of the chi-square being $\chi^2_{\rm min}=99.6$ in both cases. As before, leaving the $^{37}$Ar component free to vary does not impact significantly the results. Moreover, we foresee that in the future this should be even less problematic given that $^{37}$Ar has a half-life of about 35 days and thus should be not present in future LZ data samples.\\

\section{Conclusions}
\label{sec:conclusions}
In this paper, we describe the search for a possible neutrino electromagnetic interaction by exploiting elastic solar neutrino-electron scattering data provided by the LUX-ZEPLIN Collaboration. 
By using 331.65 t$\cdot$days of data we searched for effects of the neutrino magnetic moment and neutrino millicharge by looking for distortions in the
shape and normalization of the electron recoil spectrum. At 90\% C.L. we obtain a competitive upper limit on the effective neutrino magnetic moment, namely $\mu_{\nu}^{\rm{eff}} <  1.1 \times 10^{-11} \, \mu_{\text{B}}$, which is second only to the recent XENONnT limit. We also determined the limits considering the three different neutrino flavors separately, so the results obtained in this work can be easily compared also with experiments not sensitive to solar neutrinos. To fully exploit the potentiality of the LZ data, we also derived intriguing constraints on the neutrino millicharge discussing the impact of different interaction models, namely the FEA and the more robust and reliable EPA, showing that the EPA approach leads to much more stringent constraints. Using EPA, we obtain the current best limit on the effective neutrino millicharge $|q_{\nu}^{\rm{eff}}| <  1.5 \times 10^{-13}\;e_0$, improving significantly with respect to previous bounds. Also in this scenario, we derived the limits considering the three different flavor components, achieving also in this case the current best limits.\\

\begin{acknowledgements}
The authors wish to thank Alden Fan, Hugh Lippincott, Gregory Rischbieter, and Matthew Szydagis from the LUX-ZEPLIN Collaboration for valuable discussions and for providing crucial information on how to derive the LZ electron recoil efficiency using the NEST software.
\end{acknowledgements}

\appendix
\section{THE $Z_{\text{eff}}^{\mathcal{A}}(T_{e})$ TERM}
\label{app:zeff}

The $Z_{\text{eff}}^{\mathcal{A}}(T_{e})$ term~\cite{Mikaelyan:2002nv,Fayans:2000ns}, which quantifies the number of electrons that can be ionized by a certain energy deposit $T_e$, is given for xenon in Table~\ref{tab:electroneffcharge}. It has been obtained by using the edge energies extracted from photoabsorption data~\cite{HENKE1993181}. \\

\begin{table}[ht]
\renewcommand{\arraystretch}{1.45}
    \centering
\begin{tabular}{c|lll}

\multirow{18}{*}{$Z_{\rm eff}^{\rm Xe}$=} & 54, & & $T_{e} >$ 34.561 keV          \\
                                 & 52, & & 34.561 keV\ $\geq T_{e} >$5.4528 keV \\
                                 & 50, & & 5.4528 keV\ $\geq T_{e} >$5.1037 keV \\
                                 & 48, & & 5.1037 keV\ $\geq T_{e} >$4.7822 keV  \\
                                 & 44, & & 4.7822 keV\ $\geq T_{e} >$1.1487 keV  \\
                                 & 42, & & 1.1487 keV\ $\geq T_{e} >$1.0021 keV \\
                                 & 40, & & 1.0021 keV\ $\geq T_{e} >$0.9406 keV \\
                                 & 36, & & 0.9406 keV\ $\geq T_{e} >$0.689 keV \\
                                 & 32, & & 0.689 keV\ $\geq T_{e} >$0.6764 keV \\ 
                                 & 26, & & 0.6764 keV\ $\geq T_{e} >$0.2132 keV \\
                                 & 24, & & 0.2132 keV\ $\geq T_{e} >$0.1467 keV \\
                                 & 22, & & 0.1467 keV\ $\geq T_{e} >$0.1455 keV \\ 
                                 & 18, & & 0.1455 keV\ $\geq T_{e} >$0.0695 keV \\
                                 & 14, & & 0.0695 keV\ $\geq T_{e} >$0.0675 keV \\
                                 & 10, & & 0.0675 keV\ $\geq T_{e} >$0.0233 keV \\
                                 & 4, & & 0.0233 keV\ $\geq T_{e} >$0.0134 keV \\
                                 & 2, & & 0.0134 keV\ $\geq T_{e} >$0.0121 keV \\
                                 & 0, & & $T_{e} \leq$0.0121 keV \\  
\end{tabular}
\caption{The effective electron charge of the target atom, $Z_{\text{eff}}^{\mathrm{Xe}}(T_{e})$.}
\label{tab:electroneffcharge}
\end{table}

\section{NEUTRINO-ELECTRON COUPLING DETERMINATION}
\label{app:rad}

In order to study the neutrino-electron scattering process, it
is necessary to study in detail the calculation of the couplings, taking into account the radiative corrections.
The latter are implemented following the formalism given in Ref.~\cite{Erler:2013xha}. In particular, the $\ell$ flavor neutrino right and left couplings to fermions, with $f=e$, are given by 

\begin{align}
\label{gLLnud} 
&g_{LL}^{\nu_\ell f} = \rho \left[ - {1 \over 2} - Q_f \hat s_0^2 + \boxtimes_{ZZ}^{f L} \right] - Q_f \diameter_{\nu_\ell W} +  \Box_{WW}
\end{align}
\begin{align}
\label{gLRnu}
 g_{LR}^{\nu_\ell f} = - \rho \left[ Q_f \hat s_0^2 + \boxtimes_{ZZ}^{f R} \right] - Q_f \diameter_{\nu_\ell W} 
\end{align}
In these relations, $\rho= 1.00063$ represents a low-energy correction for neutral-current processes and $Q_f$ is the fermion charge. Here $\hat{s}_0^2=\sin^2{\vartheta}_W^{\mathrm{SM}}$, which keeps the same value for $\mu< \mathcal{O}(0.1~\mathrm{GeV})$. The other corrections inserted come from different contributions, such as the charge radii ($\diameter_{\nu_\ell W}$), and EW box diagrams ($\boxtimes_{ZZ}^{f X}$, $\square_{WW}$).
They can be expressed as
\begin{align}
\label{WWbox}
& \diameter_{\nu_\ell W} = - {\alpha \over 6 \pi} \left( \ln {M_W^2 \over m_\ell^2} + {3 \over 2} \right),\\
&\Box_{WW} = - {\hat\alpha_Z \over 2 \pi \hat s_Z^2}\left[ 1 - {\hat\alpha_s(M_W) \over 2 \pi} \right], 
\end{align}
\be\label{eq:ZZbox}
\boxtimes_{ZZ}^{fX} = - {3 \hat\alpha_Z \over 8 \pi \hat s_Z^2 \hat c_Z^2} (g_{LX}^{\nu_\ell f})^2
\left[ 1 - {\hat\alpha_s(M_Z) \over\pi} \right],
\ee
where $X\in\{L,R\}$ and 
$\hat{\alpha}_Z\equiv \alpha(M_Z)$.
Note that in Eq.~(\ref{eq:ZZbox}) all the $(g_{LX})^{\nu_\ell f}$ are evaluated at lowest order but replacing $\hat{s}^2_0$ by $\hat{s}^2_Z$ and are given by $g_{LL}^{\nu_\ell e} =-\frac{1}{2}+\hat{s}^2_Z$ and $g_{LR}^{\nu_\ell e} = \hat{s}^2_Z$.
For neutrino-electron scattering the couplings are given by
\begin{align}\nonumber
&g_V^{\nu_\ell\,e}=\rho\left(-\frac{1}{2}+2\hat{s}^2_0\right)+\Box_{WW}+2\diameter_{\nu_\ell W}\\
&+\rho(\boxtimes_{ZZ}^{eL}-\boxtimes_{ZZ}^{eR}),\\
& g_A^{\nu_\ell\,e}=\rho\left(-\frac{1}{2}+\boxtimes_{ZZ}^{eL}+\boxtimes_{ZZ}^{eR}\right)+\Box_{WW},
\end{align}
where $g_A^{\nu_\ell\,e}=g_{LL}^{\nu_\ell\,e}-g_{LR}^{\nu_\ell\,e}$.\\

For the numerical SM evaluation we assume the values from Refs.~\cite{ParticleDataGroup:2020ssz,alitti}, namely
$\hat{s}_0^2=0.23857$, 
$\hat{s}_Z^2=0.23121$,
$\alpha_s(M_W)=0.123$,
$\alpha_s(M_Z)=0.1185$, and
$\hat{\alpha}_Z^{-1}=127.952$.
We thus obtain the couplings
$g_{V}^{\nu_{e}}=0.9521$, $g_{A}^{\nu_{e}}=0.4938$, $g_{V}^{\nu_{\mu}}=-0.0397$, $g_{A}^{\nu_{\mu,\tau}}=-0.5062$, and $g_{V}^{\nu_{\tau}}=-0.0353$
that take into account all radiative corrections. We note that, for the $\nu_{e}$ coupling, an unity factor has been added to the result in order to take into account the charge current contribution.

\bibliography{ref}

\end{document}

%% file: tab_limits.tex
{\renewcommand{\arraystretch}{1.2} 
\begin{tabular}{l|c|c|c}
& \multicolumn{1}{c|}{$|\mu _{\nu }|[\times 10^{-11}\mu _{B}]$} & \multicolumn{2}{c}{$q_{\nu}\ [\times 10^{-13}e_0]$} \\
& \multicolumn{1}{c|}{} & FEA & EPA 
\\ \hline 
$\nu_{\rm eff}\ $ & $ <1.1 $ & [-3.0, 4.7] & [-1.5, 1.5]\\ \hline
$\nu _{e}\ $ & $ <1.5 $ & [-3.6, 6.5] & [-2.1, 2.0]\\ 
$\nu _{\mu }$ & $ <2.3 $ & [-8.9, 8.8] & [-3.1, 3.1]\\ 
$\nu _{\tau }$ & $ <2.1 $ & [-8.1, 8.1] & [-2.8, 2.8]\\

\end{tabular}
}